\documentclass[preprint,prd]{revtex4-1}
\begin{document}

\def\ba{\begin{eqnarray}}
\def\ea{\end{eqnarray}}
\def\w{\wedge}

%\preprint{APS/123-QED}

\title{Non-minimally Coupled Gravitational and
Electromagnetic Fields: pp-Wave Solutions}

\author{Tekin Dereli}
\email{tdereli@ku.edu.tr}
 \affiliation{Department of Physics,
Ko\c{c}
University, 34450 Sar{\i}yer, \.{I}stanbul, Turkey}
\author{\"{O}zcan Sert}
 \email{sertoz@itu.edu.tr}
\affiliation{Department of Physics Engineering, \.{I}stanbul Technical University, 34469 Maslak, \.{I}stanbul, Turkey}
 
%\date{     }

\begin{abstract}
 \noindent   We give the Lagrangian  formulation of  a generic 
non-minimally extended Einstein-Maxwell theory  with an action that is linear in the
curvature and quadratic in the electromagnetic field.  We derive the coupled field equations
by a  first order variational principle using the method of Lagrange multipliers. 
We look for solutions describing  plane-fronted Einstein-Maxwell waves with parallel rays.  
We give a family of exact pp-wave solutions  
 associated with  a partially massless spin-2 photon and a  partially massive spin-2 graviton.

\vskip 2cm

\end{abstract}
\maketitle

%%%%%%%%%%%%%%%%%%%%%%%%%%%%%%%%%%%%%%%%%%%%%%%%%%%%%%%%%%

%%%%%%%%%%%%%%%%%%%%%%%%%%%%%%%%%%%%%%%%%%%%%%%%%%%%%%%%%%%%%%%%%%%%%%%%%%%%%
\section{Introduction}

\noindent  The predictions of the classical
laws of electrodynamics have been verified to high levels of
accuracy. These are the laws that are usually  extrapolated to
describe astrophysical phenomena under extreme  conditions of
temperature, pressure and density. Any departures from these laws
under such extreme conditions  may be ascribed to new types of
interactions between the electromagnetic fields and gravity. Here we
consider non-minimal couplings of gravitational and electromagnetic fields 
 described by a Lagrangian density
that involves  generic $RF^2$-terms. 
Such a  coupling term was first considered
by Prasanna \cite{prasanna} .  They  were soon extended  and classified  by Horndeski \cite{horndeski} to gain more insight into the
relationship between space-time curvature and electric charge conservation.
 It is remarkable that a  calculation   in QED  of  the photon effective action from 1-loop vacuum polarization 
on a curved background \cite{drummond}  contributed  similar  non-minimal coupling terms.
It was  contemplated at about the same times that Kaluza-Klein reduction  of a five-dimensional $R^2$-Lagrangian 
would induce similar non-minimal couplings in four dimensions \cite{buchdahl}.
A variation of an arbitrary Lagrangian  with
non-minimally coupled gravitational and electromagnetic fields
in general may  involve field equations of order higher than two. 
The non-minimal couplings in four dimensions classified by Horndeski are exactly those that involve at most  second order terms.
These particular combinations are  obtained by  reduction of the Euler-Poincar\'{e} Lagrangian in  five dimensions to four dimensions \cite{muller-hoissen}, \cite{dereli1}.
More recently,  a 3-parameter family of
non-minimally coupled Einstein-Maxwell  field equations were  considered in various aspects in a series of papers by Balakin and Lemos \cite{balakin1},
\cite{balakin2} ,
\cite{balakin3}.  Intense gravitational fields that will  be found near  black holes behave as a specific kind of non-linear medium in the presence of  non-minimal couplings. Hence the electromagnetic waves that propagate in such  media  may imply new effects. Conversely, one should expect new gravitational effects induced by non-minimal couplings in the vicinity of neutron stars or magnetars  where there are intense electromagnetic fields. Such new effects, if there are any, can be discussed in terms of exact solutions of the coupled field equations with appropriate isometries.  

\noindent In this article, we  formulate  a 
non-minimally extended Einstein-Maxwell theory  whose Lagrangian is linear in the
curvature and quadratic in the electromagnetic field using the algebra
of exterior differential forms.  We derive the field equations
by a  first order variational principle using the method of Lagrange multipliers. The resulting system of coupled  equations we found are highly non-linear. 
Exact solutions can be obtained in  cases when the gravitational and electromagnetic fields have a high degree of symmetry. In particular,  
we consider solutions describing  plane-fronted Einstein-Maxwell waves with parallel rays  in Ehlers-Kundt  form
\cite{ehlers-kundt}, \cite{aichelburg}.  We present  a family of exact solutions  that are 
 associated with  a partially massless spin-2 photon and a  partially massive spin-2 graviton.

\section{Non-minimally
Coupled  Einstein-Maxwell Field Equations} \label{model}

\bigskip

\noindent We will derive our field equations by a variational
principle from an action
\begin{equation}
        I[e^a,{\omega^a}_b,F] = \int_M{L} = \int_M{\mathcal{L}^*1}
        \nonumber
\end{equation}
where  $\{e^a\}$ and ${\{\omega^a}_b\}$ are the fundamental
gravitational field variables and   $F$ is the electromagnetic field
2-form.  The space-time metric $g = \eta_{ab} e^a \otimes e^b$ with signature $(-+++)$ and we fix the orientation by setting $*1 = e^0 \w e^1 \w  e^2 \w e^3 $.  Torsion 2-forms $T^a$ and curvature 2-forms $R^{a}_{\; \;
b}$ of space-time are found from the Cartan-Maurer structure
equations
\begin{equation}
de^a + \omega^{a}_{\;\;b} \w e^b = T^a , \nonumber
\end{equation}
\begin{equation}
d\omega^{a}_{\;\;b} + \omega^{a}_{\;\;c} \w \omega^{c}_{\;\;b} =
R^{a}_{\;\;b} . \nonumber
\end{equation}
We consider the following Lagrangian density 4-form
 \ba
  L =  \frac{1}{2\kappa^2}R_{ab}\wedge{^*(e^a\wedge e^b)} - \frac{1}{2}F\w *F + \frac{\gamma}{2}R_{ab} \w F^{ab} *F
   \label{Lagrange} 
   \ea
where
 $\kappa^2 = 8\pi G$ is  Newton's universal gravitational constant $(c=1)$  and $ \gamma $ is a coupling
   constant. The field equations are obtained by considering the independent variations of
   the action with respect to  $\{e^a\}$,
   ${\{\omega^a}_b\}$ and $\{F\}$.  The electromagnetic field components are read  from the expansion $F = \frac{1}{2} F_{ab} e^a \w e^b$. 
We will be working with the unique metric-compatible, torsion-free
Levi-Civita connection. We impose this choice of connection through
constrained variations  by the method of Lagrange multipliers. That
is we add to the above Lagrangian density  the following constraint
terms:
\begin{equation}
L_{C} = \left ( de^a + \omega^{a}_{\;\;b} \w e^b \right ) \w
\lambda_a + dF  \w \mu  \nonumber
\end{equation}
where   $\lambda_a$'s are  Lagrange multiplier 2-forms whose
variation imposes the zero-torsion constraint  $T^a=0$. We also use a
first order variational principle for the electromagnetic field
2-forms F for which the homogeneous field equations $dF = 0$ is
imposed by the variation of the Lagrange multiplier 2-form $\mu$.

\medskip

\noindent The infinitesimal variations of the total Lagrangian density $L + L_C$
(modulo a closed form) is found to be
\begin{eqnarray}
&& \dot{L} +{\dot{L}_C} =  \frac{1}{2 \kappa^2} \dot{e}^a \w R^{bc}
\w
*e_{abc} + \frac{1}{2} \dot{e}^a \w (\iota_a F \w *F - F \w \iota_a
*F)  + \dot{e}^a \w  D \lambda_a \nonumber \\  && - \frac{\gamma}{4} \dot{e}^a \w \left ( \iota_a R_{bc} \w F^{bc} *F -
R_{bc} \w F^{bc}  \iota_a *F +  F^{bc} \iota_a F \w *R_{bc} - F^{bc}
F \w \iota_a *R_{bc}  \right )  \nonumber
\\ & &
  + \dot{\lambda}_a \w T^a   + \frac{1}{2}
\dot{\omega}_{ab} \w  ( e^b \w \lambda^a - e^a \w \lambda^b)
+\frac{\gamma}{2} \dot{\omega}_{ab} \w  D(F^{ab} *F)  \nonumber \\
& & -\dot{ F} \w
*F + \frac{\gamma}{2} \dot{F} \w F^{ab} *R_{ab} + \frac{\gamma}{2} F \w
\dot{F}^{ab} *R_{ab} - \dot{F} \w d\mu 
\end{eqnarray}
where  ${}^.$ over a field variable denotes infinitesimal variations and 
we use shorthand notation $ e^a \wedge e^b \wedge \cdots =
e^{ab\cdots}$. Lagrange multiplier 2-forms $\lambda_a$ are solved
from the connection variation equations
\ba
e_a \w \lambda_b - e_b \w \lambda_a = \gamma D(F_{ab} *F) .
\ea
It turns out that
\ba
\lambda_a = \gamma  \; \iota^b D(F_{ba}*F) - \frac{\gamma}{4} e_a \w
\iota^c \iota^b D(F_{bc} *F) .
\ea

\noindent Thus the Einstein field equations are 
\begin{eqnarray}
&& -\frac{1}{2 \kappa^2} R^{bc} \w
*e_{abc} \quad = \quad  \frac{1}{2} (\iota_a F \w *F - F \w \iota_a
*F)  +   \nonumber \\  && - \frac{\gamma}{4} \left ( \iota_a R_{bc} \w F^{bc} *F -
R_{bc} \w F^{bc}  \iota_a *F +  F^{bc} \iota_a F \w *R_{bc} - F^{bc}
F \w \iota_a*R_{bc}  \right ) \nonumber \\
&& + \gamma F_{ac} \iota_bF \w *R^{cb} + \gamma D(\iota^b
D(F_{ba}*F)) - \frac{\gamma}{4} e_a \w D(\iota^c \iota^b D(F_{bc}
*F))  
\end{eqnarray}
while the Maxwell equations read
\begin{equation}
dF = 0 \quad , \quad d*(F - \gamma F_{ab} R^{ab}) = 0 . 
\end{equation}

\subsection{Electromagnetic
Constitutive Equations}

\noindent In general one may encode the effects of non-minimal
couplings of the electromagnetic fields to gravity into the
definition of a constitutive tensor. Maxwell 's equations for an
electromagnetic field $F$ in an arbitrary medium can be written as
\ba
dF = 0 \quad , \quad *d*G = J
\ea
where $G$ is called the excitation 2-form and $J$ is the source
electric current density 1-form. The effects of gravitation and
electromagnetism on matter are described by $G$ and $J$. To close
this system we need electromagnetic constitutive relations relating
$G$ and $J$ to $F$.  Here we consider only the source-free
interactions, so that $J=0$. Then we take a simple linear
constitutive relation
\ba
G = {\cal{Z}} (F)
\ea
where ${\cal{Z}}$ is a type-(2,2)-constitutive tensor. For the above
theory we have
\ba
G = F - \gamma R_{ab}  F^{ab} .
\ea
With these definitions,  the non-minimally coupled Einstein-Maxwell
action density simply becomes
\ba
   L =  \frac{1}{2\kappa^2}\mathcal{R } *1 - \frac{1}{2}F\w *G .
\ea
The electric field $\emph{e}$ and magnetic induction field
$\emph{b}$ associated with $F$ are defined with respect to an
arbitrary unit future-pointing time-like 4-velocity vector field $U$
("inertial observer") by
\ba
\emph{e} = \iota_{U} F  \quad , \quad\emph{ b} = \iota_{U} *F .
\ea
Since $g(U,U) = -1$ we have
\ba
F = \emph{e} \w \tilde{U} - *(\emph{b} \w \tilde{U}) .
\ea
Likewise the electric displacement field $\emph{d}$ and the magnetic
field $\emph{h}$ associated with $G$ are defined with respect to $U$
as
\ba
\emph{d} = \iota_{U} G  \quad , \quad\emph{ h }= \iota_{U} *G .
\ea
Thus
\ba
G = \emph{d} \w \tilde{U} - *(\emph{h} \w \tilde{U}) .
\ea
It is sometimes convenient to work in terms of polarization 1-form
$\emph{p} \equiv \emph{d} - \emph{e} = - \gamma ( \iota_{U}R_{ab}) F^{ab}$ and magnetization 1-form
$\emph{ m} \equiv \emph{b} - \emph{h} =  \gamma ( \iota_{U}*R_{ab}) F^{ab}. $

 \section{ Plane-Fronted Wave Solutions}

\noindent We seek solutions that describe plane fronted waves with
parallel rays (pp-waves). A generic pp-wave metric in Ehlers-Kundt
form \cite{ehlers-kundt}  is given by,
\begin{equation}
              g = 2dudv  +  dx^2 + dy^2 + 2H(u,x,y)du^2 .
\end{equation}
H is a smooth function to be determined. A convenient choice of
orthonormal co-frames is going to be used:
\begin{equation}
           e^0 = \frac{H-1}{\sqrt{2}}du + dv, \quad   e^1 = dx, \quad e^2 =
           dy, \quad
            e^3 = \frac{H+1}{\sqrt{2}}du + dv.
\end{equation}
We may also exploit the advantages of complex coordinates in
transverse plane  by letting
\begin{equation}
              g = 2dudv  +  2dzd\overline{z} +2H(u,z,\overline{z})du^2
\end{equation}
 where $$z=\frac{x+i y}{\sqrt{2}} \quad , \quad \bar{z}=\frac{x-i y}{\sqrt{2}} .$$ The non-vanishing Levi-Civita
 connection 1-forms are
\begin{equation}\label{w_ab}
             {\omega^0}_1 = -{\omega^1}_3 = \frac{H_x}{2}(e^3-e^0),
             \quad
             {\omega^0}_2 = - {\omega^2}_3 = \frac{H_y}{2}(e^3-e^0).
             \nonumber
\end{equation}
Then the  Einstein 3-forms  $  G_a = -\frac{1}{2} R^{bc} \w
*e_{abc}$ become
 \begin{eqnarray}
  G_0 = -G_3 =  \frac{H_{xx} + H_{yy}}{2}*(e^3-e^0), \quad
G_1 = 0 = G_2 . \nonumber
\end{eqnarray}
We consider an electromagnetic potential 1-form given as $A =
a(u,x,y) du $ or $A=a(u,z,\overline{z}) du$ for pp-waves. Then
 \begin{eqnarray}
 F  &=&  dA \nonumber\\
   &=& a_x dx \w du + a_y  dy \w du \nonumber \\ &=&  a_z dz \w du + a_{\overline{z}}d\overline{z} \w
   du 
\end{eqnarray}
We substitute these into $(9)$ and $(10)$  and after a lengthy calculation reach the final form of the  the non-minimally
coupled Einstein-Maxwell equations 
\begin{eqnarray}
H_{xx} + H_{yy} & = & -\kappa^2 ({a_x}^2 +
 {a_y}^2) + \kappa^2 \gamma \left ( ({a_x}^2)_{xx}
+ 2( a_x a_y )_{xy}   +  ({a_y}^2)_{yy} \right ) , \nonumber \\
a_{xx} + a_{yy} &=& 0 . 
\end{eqnarray}
These equations can be re-written in an invariant form on the
transverse $xy$-plane \cite{aichelburg},\cite{dereli2}:
\begin{eqnarray}
\Delta H &=& -\kappa^2 |\nabla a|^2 + \kappa^2 \gamma Hess(a)
\nonumber \\ & & - 2 \kappa^2 \gamma \left (  \Delta(a \Delta a) - a \Delta (\Delta a) + (\Delta a)^2  \right ) \nonumber \\
\Delta a & = & 0 
\end{eqnarray}
where $$ \Delta = \frac{\partial^2}{\partial x^2} +
\frac{\partial^2}{\partial y^2}$$ is the 2-dimensional Laplacian,
$$|\nabla a|^2 = \left ( \frac{\partial a}{\partial x} \right )^2 +
\left ( \frac{\partial a}{\partial y} \right )^2$$ is the
norm-squared of the 2-dimensional gradient and
$$
Hess(a) = \left | \begin{array}{cc} a_{xx} &  a_{xy} \\ a_{yx} &
a_{yy}\end{array} \right | = a_{xx} a_{yy} - (a_{xy})^2
$$ is the 2-dimensional Hessian operator.
In terms of complex coordinates,  $(25)$ simply
reads
\ba
H_{z \bar{z}}  = -\kappa^2 a_{z} a_{\bar{z}} + \kappa^2 \gamma a_{zz}
a_{\bar{z}\bar{z}}  \quad , \quad
a_{z \bar{z}}  = 0 . 
\ea
A non-trivial solution that depends on the coupling constant
$\gamma$ is obtained by letting
\ba
   a(u,z,\overline{z})= f_1(u)z + \bar{f}_1(u)\overline{z} + f_2(u) z^2 + \bar{f}_2(u)
   \overline{z}^2 .
\ea
Then
\begin{eqnarray}\label{complexdenk}
\frac{1}{\kappa^2} H(u,z,\overline{z}) &=&   f_3(u) z^2 +
\bar{f}_3(u) \bar{z}^2 -  |f_1(u)|^2|z|^2 - |f_2(u)|^2 |z|^4
\nonumber \\ & & - f_1(u) \overline{f}_2(u) \bar{z} |z|^2 - f_2(u)
\overline{f}_1(u) z |z|^2  + 4 \gamma |f_2|^2 |z|^2  .
\end{eqnarray}

%%%%%%%%%%

\noindent  We  note that the non-minimal coupling $\gamma$  between the gravitational and electromagnetic waves is carried in the 
last term on the right hand side of the expression above and affects only the space-time metric.  
Both the polarization $p = 0$ and the magnetization $m = 0$     identically in the pp-wave geometry. 
We write
\ba
A = \mathcal{A}_{1+}  + \mathcal{A}_{1-} +  \mathcal{A}_{2+}  + \mathcal{A}_{2-} 
\ea
where
\ba
\mathcal{A}_{1+}  =   f_1(u) z   du = \bar{\mathcal{A}}_{1-}  \quad , \quad
\mathcal{A}_{2+}  =   f_2(u) z^2   du = \bar{\mathcal{A}}_{2-} 
\ea
and  introduce $z = re^{i\theta}$   to show that 
\begin{eqnarray}
 \mathcal{L}_{ \frac{1}{i} \frac{\partial }{ \partial \theta} } \mathcal{A}_{1 \pm} =  \pm \mathcal{A}_{1 \pm}
 \hskip 1 cm  \mathcal{L}_{\frac{1}{i} \frac{\partial }{ \partial \theta}  } \mathcal{A}_{2 \pm }
 =
 \pm  2 \mathcal{A}_{2 \pm} .
  \end{eqnarray}
  $\mathcal{L}_X  $ denotes the Lie derivative along the
  vector field $X$. Hence $\mathcal{A}_{1 \pm} , \mathcal{A}_{2\pm}$ are null photon helicity eigen-tensors. 
  Similarly, 
the metric tensor decomposes as
\begin{eqnarray}
g= \eta + \mathcal{G}_{0} + \mathcal{G}_{1+} +  \mathcal{G}_{1-} + \mathcal{G}_{2+}
+ \mathcal{G}_{2-} 
  \end{eqnarray}
  where $\eta$ is the metric of Minkowski spacetime and
\begin{eqnarray}
 \mathcal{G}_{1+} &=& - \bar{f}_1(u) f_2(u)z |z|^2 du\otimes du =
 \bar{\mathcal{G}}_{1-}\\
  \mathcal{G}_{2+} &=& \bar{f}_3(u) z^2  du\otimes du = \bar{\mathcal{G}}_{2-}
  \\
   \mathcal{G}_{0} &=& \left ( -|{f}_1(u)|^2  -|f_2(u)|^2 |z|^4+4\gamma |f_2(u)|^2|z|^2 \right ) du\otimes
   du.
  \end{eqnarray}
The $  \mathcal{G}_{1\pm},   \mathcal{G}_{2\pm}$ are null g-wave
helicity eigen-tensors for linearized gravitation about $\eta +
\mathcal{G}_{0}$:
\begin{eqnarray}
 \mathcal{L}_{ \frac{1}{i} \frac{\partial }{ \partial \theta} } \mathcal{G}_{1 \pm} =  \pm \mathcal{G}_{1 \pm}
 \hskip 1 cm  \mathcal{L}_{\frac{1}{i} \frac{\partial }{ \partial \theta}  } \mathcal{G}_{2 \pm }
 =
 \pm  2\mathcal{G}_{2 \pm} .
  \end{eqnarray}
 This is a configuration associated with  a partially massless spin-2 photon \cite{deser}, \cite{zinoviev} and a partially massive spin-2 graviton.

%%%%%%%%%%%%%%%%%%%%%%%%%%%%%%%%%%%%%%%%
\section{ Conclusion}

\noindent We have derived the  field equations of a non-minimally coupled Einstein-Maxwell
theory by a first order variational principle using the
method of Lagrange multipliers in the language of exterior differential forms.  We give a  class of exact, non-trivial pp-wave
solutions. These solutions describe parallely propagating plane fronted gravitational and electromagnetic waves that do not interact with each other in the Einstein-Maxwell theory. 
Here if only the standard degrees of polarization  ( $\pm1$ for the photon and $\pm2$ for the graviton) are kept, no contribution arises from the non-minimal coupling constant $\gamma$. 
It is interesting to note, however,  that  if  $\gamma$ is kept  it brings in $\pm2$ polarization degrees of freedom for the photon  together with 
 $\pm1$ polarization degrees of freedom for the graviton.  The notion of a partially massless (spin-2) photon had been introduced before by Deser and Waldron \cite{deser}, \cite{zinoviev}
On the other hand, the partially massive (spin-2) graviton here is new and it may find some observational evidence in future. 
We wish to conclude by a few comments:
\begin{itemize}
\item  It is possible to  write an arbitrary linear combination of all $RF^2$-type
invariants \cite{horndeski} as  additional terms in  the Lagrangian density 4-form 
\begin{eqnarray}
L^{\prime}  &=&
      \frac{\gamma_2}{2 }\imath^a F \wedge  \mathcal{R}_a \wedge *F
     +\frac{\gamma_3}{2 }  \mathcal{R} F\wedge *F \nonumber \\
     &&
+   \frac{\gamma_4}{2 } {R}_{ab}F^{ab}\wedge F
 +     \frac{\gamma_5}{2 }\imath^a F \wedge \mathcal{R}_a \wedge F
  +  \frac{\gamma_6}{2 }  \mathcal{R} F\wedge F 
\end{eqnarray}
where  $\mathcal{R}_a = \iota_b {R}^{b}_{\; \; a}$
are the Ricci 1-forms and $\mathcal{R}$ is the curvature scalar. 
Now the variational  field equations become a
lot more complicated but no essential insight
is gained by such a generalization as far as the pp-wave solutions above are concerned.  

\item A conformally scale invariant non-minimal coupling is achieved (for the case $\omega = -\frac{3}{2}$) 
by considering
\ba
L = \frac{\phi}{2} R_{ab} \w *e^{ab} - \frac{\omega}{2 \phi} d \phi
\w *d\phi - \frac{1}{2} F \w *F + \frac{\gamma}{2 \phi} C_{ab} \w
F^{ab} *F
\ea
 where $\phi$ is the dilaton and  
\ba
C_{ab} = R_{ab} - \frac{1}{2} ( e_a \w \mathcal{R}_b - e_b \w \mathcal{R}_a ) +
\frac{1}{6}{\mathcal{R}} e_{ab} 
\ea
are the Weyl conformal curvature 2-forms  $(  \gamma_2 = \gamma, \gamma_3 = \frac{\gamma}{3} , \gamma_4 = \gamma_5 = \gamma_6 = 0 )$ .

\item One may give up the zero-torsion constraint and vary the
action   $(4)$ with respect to the metric and connection treated as independent
variables.  The equations resulting from the connection variations can be solved for the torsion 2-forms
\ba
T_a = \kappa^2 \gamma \; \iota^b D(\tilde{F}_{ab} *F) + 
\frac{1}{4}\kappa^2 \gamma \; e_a \w \iota^c \iota^b
D(\tilde{F}_{bc}
*F) . \label{torsion}
\ea
Here we use  the expansion $*F = \frac{1}{2} \tilde{F}_{ab} e^{ab}$.  The exterior covariant derivatives on the right hand side implicitlly involve the contortion 1-forms. 
Therefore, the algebraic equations (\ref{torsion}) admit a unique solution for the torsion 2-forms in terms of the tensor $\hat{D}(\tilde{F}_{ab} *F)$,
but an explicit formula is not easy to obtain.

\end{itemize}

\vskip 1cm

\begin{acknowledgements}

\noindent  T.D. gratefully acknowledges partial  support from The Turkish Academy of Sciences (TUBA).  
The research of \"{O}.S. is supported by a grant from the Scientific and Technological Research Council of Turkey (TUBITAK). 

\end{acknowledgements}

\vskip 1cm

%\newpage


\begin{thebibliography}{99}

\bibitem{prasanna}  A. R. Prasanna,  {\it Phys. Lett.}  {\bf A37}, 331 (1971) .

\bibitem{horndeski}  G. W. Horndeski, {\it J. Math. Phys.} {\bf 17}, 1980 (1976).

\bibitem{drummond}  I. T. Drummond, S. J. Hathrell, {\it Phys. Rev. } {\bf D22}, 343
(1980).

\bibitem{buchdahl} H. A. Buchdahl, {\it J. Phys} {\bf A12}, 1037 (1979).

\bibitem{muller-hoissen} F.  M\"{u}ller-Hoissen, {\it Class. Q. Grav.} {\bf 5}, L35 (1988).

\bibitem{dereli1}  T. Dereli, G. \"{U}\c{c}oluk, {\it Class. Q. Grav.} {\bf 7},
1109 (1990) .

\bibitem{balakin1}  A. B. Balakin, J.P.S. Lemos, {\it Class. Q. Grav.} {\bf 18}, 941 (2001).

\bibitem{balakin2}  A. B. Balakin,  J. P. S. Lemos,  {\it Class. Q. Grav.}  {\bf 19}, 4897 (2002).

\bibitem{balakin3}  A. B. Balakin,  J. P. S. Lemos,  {\it Class. Q. Grav.}  {\bf 22}, 1867 (2005).

\bibitem{ehlers-kundt}  J. Ehlers, W. Kundt in {\it Gravitation: An Introduction to
Current Research}, edited by L. Witten (Wiley, New York,  1962).

\bibitem {aichelburg}  P. C. Aichelburg  {\it Acta Phys. Austr.}  {\bf 34}, 279 (1971).

\bibitem{dereli2} T. Dereli, R. W. Tucker {\it Class. Q. Grav.}  {\bf 21 } ,
1459 (2004) .

\bibitem{deser} S. Deser,  A. Waldron, {\it Phys. Rev. } {\bf D74}, 084036 (2006). 

\bibitem{zinoviev}  Yu. M. Zinoviev, Nucl. Phys. {\bf B821}, 431 (2009). 

\end{thebibliography}
\end{document}